\documentclass[12pt, twocolumn]{openjournal}
\usepackage{CJK}
\usepackage{hyperref}
\hypersetup{
    colorlinks=true,
    linkcolor=blue,
    filecolor=blue,      
    urlcolor=blue,
    citecolor=blue,
}
\usepackage{orcidlink}
\usepackage{float}

\def\dim#1{{\rm\,#1}}
\def\Lya{Ly$\alpha$}
\def\Lyb{Ly$\beta$}
\def\lya{{{\rm Ly}\alpha}}
\def\HI{{\rm H\scriptstyle\rm I}}

\shorttitle{}
\shortauthors{Gnedin \& Zhu}

\begin{document}
\begin{CJK*}{UTF8}{gkai}

\title{Damping Wing-Like Features in the Spectra of High Redshift Quasars: a Challenge for Fully-Coupled Simulations}

\email{ngnedin@gmail.com}
\author{Nickolay Y.\ Gnedin\orcidlink{0000-0001-5925-4580}}
\affiliation{Theory Division; 
Fermi National Accelerator Laboratory;
Batavia, IL 60510, USA}
\affiliation{Kavli Institute for Cosmological Physics;
The University of Chicago;
Chicago, IL 60637, USA}
\affiliation{Department of Astronomy \& Astrophysics; 
The University of Chicago; 
Chicago, IL 60637, USA}

\author{Hanjue Zhu (朱涵珏)\orcidlink{0000-0003-0861-0922}}
\affiliation{Department of Astronomy \& Astrophysics; 
The University of Chicago; 
Chicago, IL 60637, USA}

\begin{abstract}
Recently, several observational detections of damping-wing-like features at the edges of ``dark gaps" in the spectra of distant quasars (the ``Malloy-Lidz effect") have been reported, rendering strong support for the existence of ``neutral islands" in the universe at redshifts as low as $z<5.5$. We apply the procedure from one of these works, \citet{YZhu2024}, to the outputs of fully coupled cosmological simulations from two recent large projects, ``Cosmic Reionization On Computers" (CROC) and ``Thesan". Synthetic spectra in both simulations have statistics of dark gaps similar to observations, but do not exhibit the damping wing features. Moreover, a toy model with neutral islands added ``by hand" only reproduces the observational results when the fraction of neutral islands among all dark gaps approaches 90\%. I.e., simulations and observations appear to produce two distinct ``populations" of dark gaps. In addition, in the simulations, the neutral islands at $z=5.9$ should be short-lived and should not extend to $z<5.5$. A possible explanation for this discrepancy is that both simulations underestimate the fluctuations in the photoionization rate and, hence, miss a population of long-lived neutral islands, located in the large downward fluctuations of the photoionization rate. 
\end{abstract}


\section{Introduction}\label{sec:intro}

The last stages of cosmic reionization remain a subject of intense study, with particular interest in whether substantial reservoirs of neutral hydrogen persist at redshifts below 6. Over the last decade, several lines of evidence have been put forward to support the hypothesis of these so-called “neutral islands,” each offering interesting but indirect probes into the ionization state of the intergalactic medium (IGM). The first such argument was based on the observed distribution of mean opacities in the spectra of high redshift quasars \citep{Bosman2018taus,Eilers2018taus,taus2019,taus2020,Nasir2020,Bosman2022taus} and another relies on the observed abundance of long gaps in the spectra \citep{Becker2015gaps,YZhu2021gaps}. Unfortunately, both of these arguments are model-dependent, and there exist counterexamples to both of them, albeit these counterexamples themselves are model-dependent and hence are not fully conclusive. \citet{Yang2020taus} showed that their mean neutral fraction measurements of $\langle x_\HI\rangle_V < 10^{-4}$ for $z<6$ are consistent with those of \citet{Fan2006}, requiring the end of reionization by $z\approx 6$ and the volume fraction occupied by neutral islands to be significantly below $10^{-4}$. \citet{Gnedin2022} presented two simulations that complete reionization by $z=6.7$, one matching the observed distribution of the effective optical depth and another matching the gap abundance, although no single simulation matches both. \citet{Garaldi2022} showed another numerical simulation that reionizes at $z\approx 6.5$ and matches the observed distribution of the effective optical depth, while their late ($z\sim 5.5$) reionizing models fail to match the observed distribution of the effective optical depth at $z\approx 6$.

\begin{figure*}[t]
\centering
\includegraphics[width=\textwidth]{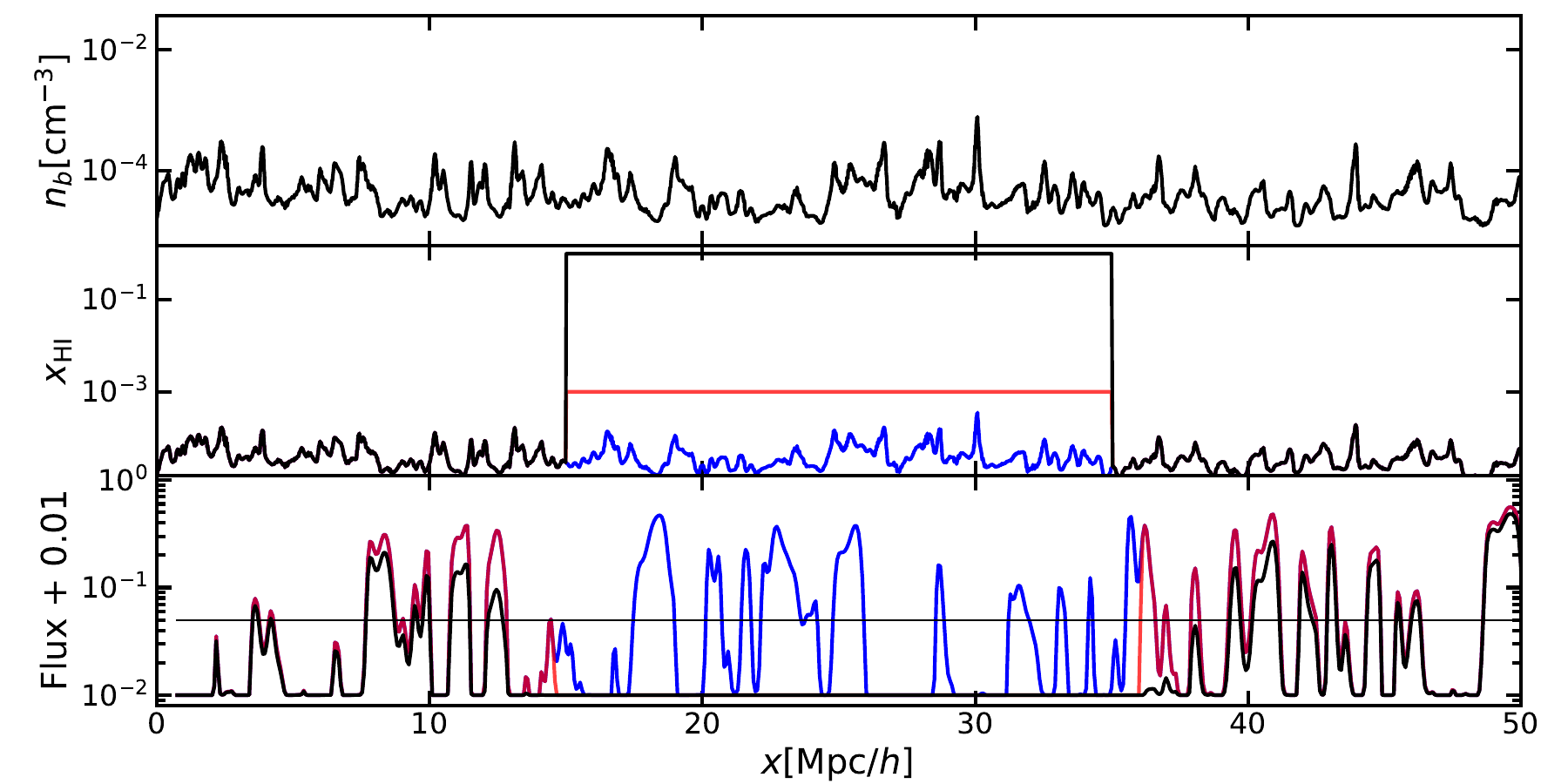}%
\caption{An illustration of the difficulty of constraining the hydrogen neutral fraction in the IGM from quasar spectra. The three panels show the baryon density, the neutral hydrogen fraction, and the synthetic quasar spectrum (in log scale to show the full dynamic range) along a line of sight from a numerical simulation at $z=5.5$ (blue). The neutral fraction is also artificially set to $10^{-3}$ and 1 in the central $20/h$ cMpc as a model for a ``neutral island" (translucent red and black). The only difference between $x_\HI=10^{-3}$ and $x_\HI=1$ is the flux suppression, seen as the difference between the black and purple (the overlay of solid blue and translucent red) lines, immediately outside the ``neutral island" due to the damping wings.}
\label{fig:island}
\end{figure*}

The challenge of using indirect probes of the ionization state of the IGM at $z > 5$ is illustrated in Figure \ref{fig:island}. It shows a synthetic spectrum at $z=5.5$ from one of the simulations described below. The original spectrum, shown in blue, exhibits well-known transmission spikes. In black, we show a toy model that introduces a large ``neutral island" by manually setting the hydrogen neutral fraction to 1 everywhere between $15/h$ and $35/h$ cMpc. The resulting synthetic spectrum has no flux within this spatial region. Finally, we also show (in translucent red) another artificial case where the neutral fraction inside the ``island" is set to $10^{-3}$ instead of 1. Since the optical depth of the neutral IGM at the mean density at this redshift is around $6\times10^5$, even in deep voids (with overdensity of a few percent), a neutral fraction of $10^{-3}$ produces complete absorption ($\tau_\lya \sim 10$). The only difference between $x_\HI=10^{-3}$ and $x_\HI=1$ is the suppression of flux \emph{outside} of a neutral region compared to the $x_\HI=10^{-3}$ case, visible as the difference between black and purple (the overlay of solid blue and translucent red) lines between $\sim5/h$ and $15/h$ cMpc and between $35/h$ and $\sim 45/h$ cMpc. This suppression arises from the extended damping wing of the \Lya\ absorption cross-section, which has a Lorentzian profile with broad wings that fall off as $(\nu - \nu_0)^{-2}$, where $\nu$ is the photon frequency and $\nu_0$ is the \Lya\ resonance frequency. As a result, even regions of the IGM that lie outside a neutral patch and are themselves highly ionized can still experience significant absorption due to the proximity of neutral hydrogen. Because the damping wing optical depth is directly proportional to the total column density of neutral hydrogen, it differs substantially between $x_{\rm HI} = 10^{-3}$ and $x_{\rm HI} = 1$. In the fully neutral case, the higher neutral hydrogen abundance leads to a strong damping wing, which absorbs photons over a broader frequency (or spatial) range beyond the edges of the region. In contrast, a lower neutral fraction produces no damping wings and thus minimal absorption outside the core of the patch.

As was first pointed out by \citet{Malloy2015}, detecting such flux suppression is possibly the only robust signature of a genuine mostly neutral region in the IGM. Recently, three different papers reported the detection of the ``Malloy-Lidz" effect in the observational data \citep{Becker2024,Spina2024,YZhu2024}. There are two main observational challenges with such detection. First, the spectra are noisy, and transmission spikes are rare, so it is difficult to detect the damping wing effect in a single spectrum \citep[but see][]{Becker2024} and hence stacking several spectra can be beneficial \citep{Spina2024,YZhu2024}. The second challenge is that the actual locations of neutral islands are not known, so from the \Lya\ absorption spectrum alone, the edge of the damping wing cannot be determined. \citet{Becker2024}, \citet{Spina2024}, and \citet{YZhu2024} solved the latter problem by cleverly using the \Lyb\ absorption spectra to determine where the gaps end. In all three papers, the detection has been claimed, with perhaps the most easily interpretable example provided by Figure 1 of \citet{YZhu2024}.

There are at least two potential pitfalls in interpreting these detections. First, \Lyb\ gaps are expected to be narrower than \Lya\ gaps. This is because transmission spikes predominantly arise in deep voids \citep{Garaldi2019,HZhu2024}, and since \Lyb\ has a lower absorption cross-section, it is more likely to show a spike where \Lya\ remains fully absorbed. As a result, observers using \Lyb\ transmission spikes to define the ends of gaps may inadvertently align the \Lya\ spectra too early - that is, while \Lya\ is still within the dark part of the gap. This can artificially create a reduction in \Lya\ flux near the nominal edge, mimicking a damping wing even in the absence of a neutral region \citep{YZhu2024}. The second pitfall goes in the opposite direction: if the \Lyb\ transmission spike that ends a gap also appears in \Lya, then stacking \Lya\ spectra at that position will reveal a flux spike, not a suppression. This could mask the presence of a genuine damping wing or lead to a mistaken conclusion that the IGM is fully ionized. In both cases, careful interpretation is required, as the behavior of \Lyb\ and \Lya\ transmission spikes is not always one-to-one, and incorrect alignment can either mimic or obscure the subtle signal of a damping wing.

\begin{figure}
\centering
\includegraphics[width=\columnwidth]{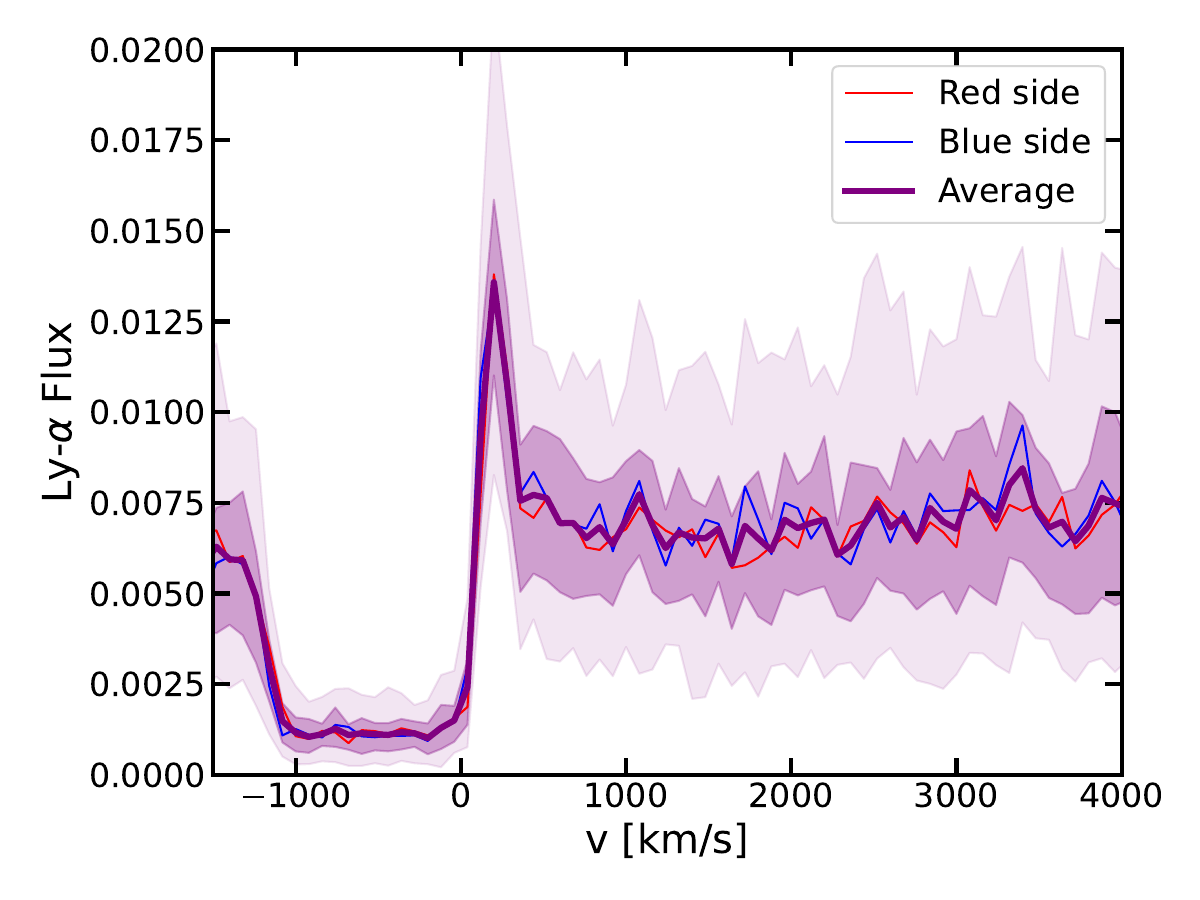}\newline
\includegraphics[width=\columnwidth]{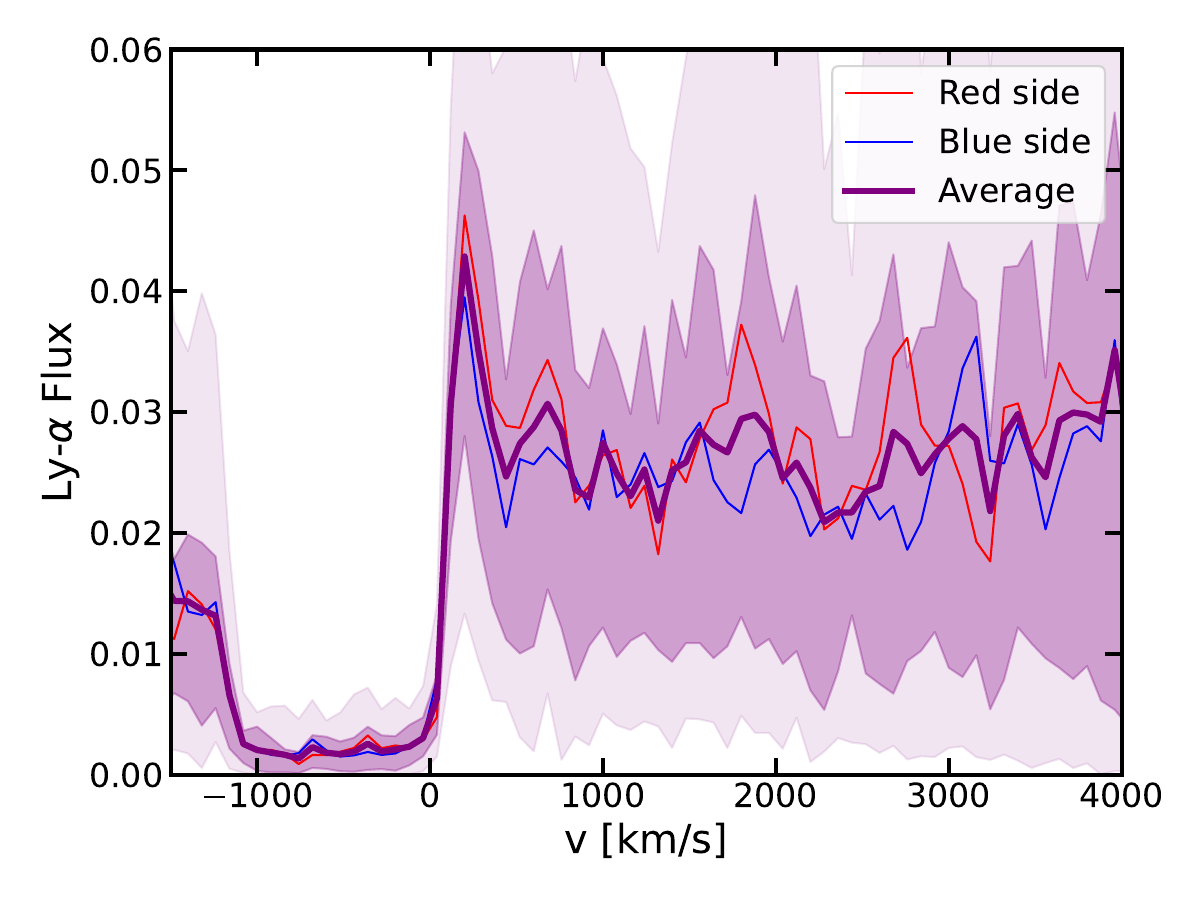}%
\caption{Stacked \Lya\ forest corresponding to the redshift of edges of \Lyb\ dark gaps (an equivalent to Figure\ 1 of \protect\citet{YZhu2024}) at $z=5.9$. The top panel shows CROC and the bottom panel shows Thesan-1 simulations. At this moment, the average volume-weighted neutral fraction is $1.5\times10^{-4}$ in CROC and 0.046 in Thesan, and the mean \Lya\ flux is the mean value outside the gaps (0.0075 for CROC and 0.25 for Thesan). Blue and red lines show flux away from the blue and red sides of the dark gap, respectively, and the thick purple line plots their average. The darker and fainter translucent bands show the 5\%-95\% and 25\%-75\% percentile ranges, respectively, for sub-samples of 24 lines of sight (the size of \protect\citet{YZhu2024} sample). No damping-wing-like feature is observed in either simulation.}
\label{fig:gaps}
\end{figure}

\begin{figure}
\centering
\includegraphics[width=\columnwidth]{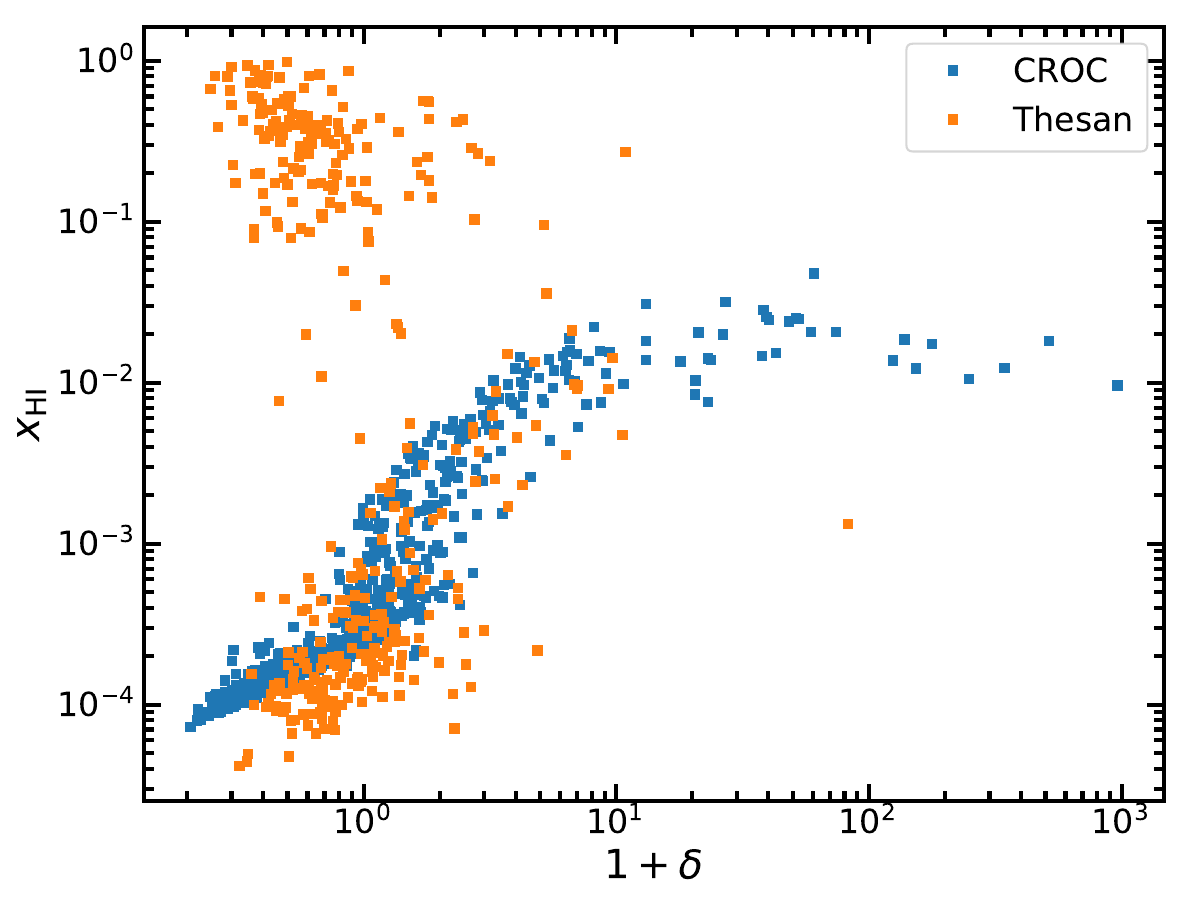}\newline
\includegraphics[width=\columnwidth]{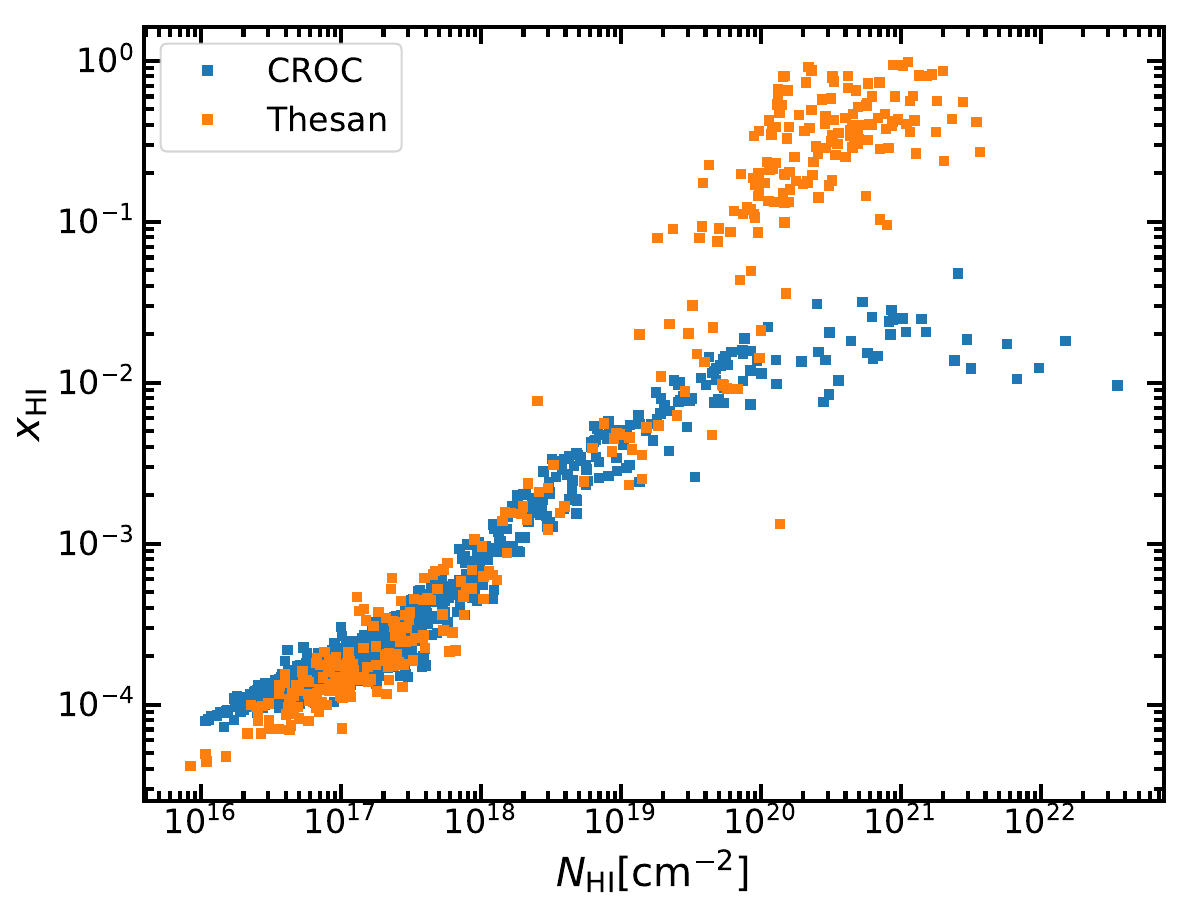}%
\caption{Properties of dark gaps (mean neutral fraction vs the mean density and the column density, averaged over the full length of each gap) in CROC and Thesan at $z=5.9$. Many dark gaps in both simulations come \emph{not} from neutral islands.}
\label{fig:props}
\end{figure}

The original goal of this short paper was to explore whether these potential pitfalls subvert the observational detections and to check whether the damping-wing-like feature at the edge of a gap is indeed associated with ``neutral islands". For that task, the locations of ``neutral islands" must be known, so this can only be done in numerical simulations of cosmic reionization. Unfortunately, as we show below, the simulations available to us fail to match the observations by a big margin, so the original goal is not achieved. Instead, the focus of the paper is now to identify discrepancies between the simulations and observations and to discuss potential reasons for the simulation failure.

The two simulation sets available to us are the ``Cosmic Reionization On Computers" \citep[CROC,][]{Gnedin2014,GnedinKaurov2014,Gnedin2022} and ``Thesan" \citep{Kannan2022,Garaldi2022,Garaldi2024}. The two simulations are very similar in many respects: they have similar volumes ($\sim 100$ cMpc) and spatial resolutions (100-300 proper pc), almost identical mass resolution ($2000^3$), and model a similar range of relevant physical effects: gas dynamics, non-equilibrium cooling and ionization, radiative transfer, and star formation and stellar feedback. They both marginally resolve all the structures in the photoionized IGM after reionization and capture the key physical effects that may be important during that time: spatial fluctuations in the photo-ionization rate, spatial and temporal variations in the gas temperature versus density, proximity effects around sources, large-scale fluctuations in the photon mean free path, etc. Hence, CROC and Thesan are some of the most suitable numerical tools available for exploring the subtle effects in quasar spectra.

\section{Results}\label{sec:results}

\begin{figure*}
\centering
\includegraphics[width=\textwidth]{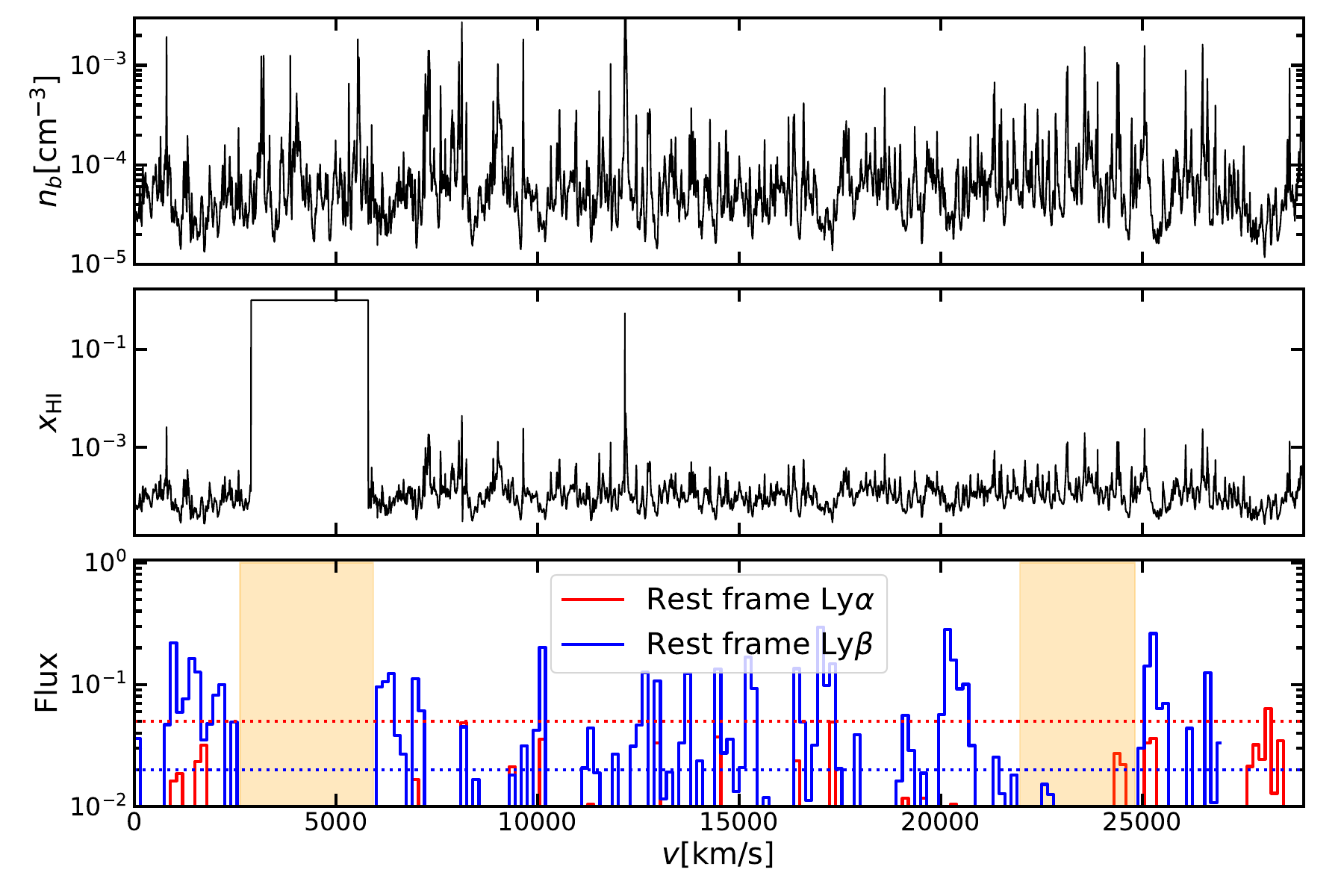}%
\caption{The three panels show the baryon density, the neutral hydrogen fraction, and the synthetic quasar spectrum (in log scale to show the full dynamic range and binned to 150 km/s resolution as in \citet{YZhu2021gaps}) along one line of sight from the CROC simulation with a neutral island added by hand between 3{,}000 and 6{,}000 km/s. Another dark gap, present in the original simulation, is seen between 22{,}000 and 25{,}000 km/s. Two horizontal dotted lines mark the \citet{YZhu2021gaps} threshold for dark gap definition: 0.05 and 0.02 for \Lya\ and \Lyb\ respectively.}
\label{fig:los}
\end{figure*}

Producing synthetic \Lyb\ spectra from simulations is non-trivial. While synthetic \Lya\ spectra for a narrow redshift interval can be easily produced from the simulation data, the \Lyb\ forest overlaps with the \Lya\ forest at a lower redshift. High-resolution simulation boxes are not large enough to cover the full spectral range from \Lya\ to \Lyb. Hence, at least two independent snapshots must be used. For the case in question, \cite{YZhu2024} identified gaps at the \Lya\ transition wavelength at $z\sim5.9$. The \Lyb\ transition at this redshift has the same wavelength as the \Lya\ transition at $z=1026\,\rm{\AA}/1216\,\rm{\AA} \times(1+5.9)-1=4.8$. CROC simulations barely reach that cosmic time (being stopped at $z=5$), and the largest Thesan simulation, Thesan-1, is stopped even earlier. In addition, different synthetic spectra are usually generated along random lines of sight at each simulation snapshot, but the random directions and starting points are often the same for all snapshots (this is the case for both CROC and Thesan). Hence, care must be taken to ensure that the \Lya\ forest at $z=5.9$ and $z=4.8$ are not artificially correlated by accidentally using identically oriented sightlines.

This is easy to achieve for CROC since it includes several simulations starting with independent realizations of cosmological initial conditions. Here we use the simulation box labeled ``C80C.DC=-1" for $z=5.9$ \Lya\ and $z=5$ \Lyb\ forests and another run, ``C80A", for the $z=5$ \Lya\ forests. The ``C80C.DC=-1" box matches the observational constraints on the abundance of dark gaps \citep{Gnedin2022} and hence is the most suitable of all CROC boxes for modeling dark gaps. For Thesan-1, we select two different lines of sight for the $z=5.9$ and $z=5$ \Lya\ forest. In addition, the publicly released Thesan-1 data do not actually include synthetic quasar spectra at $z=5$. Instead, we use the latest available snapshot at $z=5.5$ and rescale the \Lya\ forest at that snapshot to $z=4.8$ using the average evolution of the \Lya\ opacity from \citet{Yang2020taus}. The abundance of the dark gaps in Thesan-1 is similar to that in CROC ``C80C.DC=-1" run and is, therefore, approximately consistent with observations of \citet{YZhu2021gaps}. Notice, however, that both simulations do not match the observed distribution of mean opacities \citep{Bosman2022taus}.

We then follow the \citet{YZhu2024} procedure to generate synthetic observations: identify \Lya\ and \Lyb\ gaps in the spectra as continuous regions with the flux below the thresholds of 0.05 and 0.02, align the \Lyb\ gaps at their left and right edges, and compute the average \Lya\ flux outside the gap edge as a function of velocity away from the edge. We use 1000 lines of sight for CROC and 300 for Thesan (all that is available). 

\begin{figure*}[t]
\centering
\includegraphics[width=0.5\textwidth]{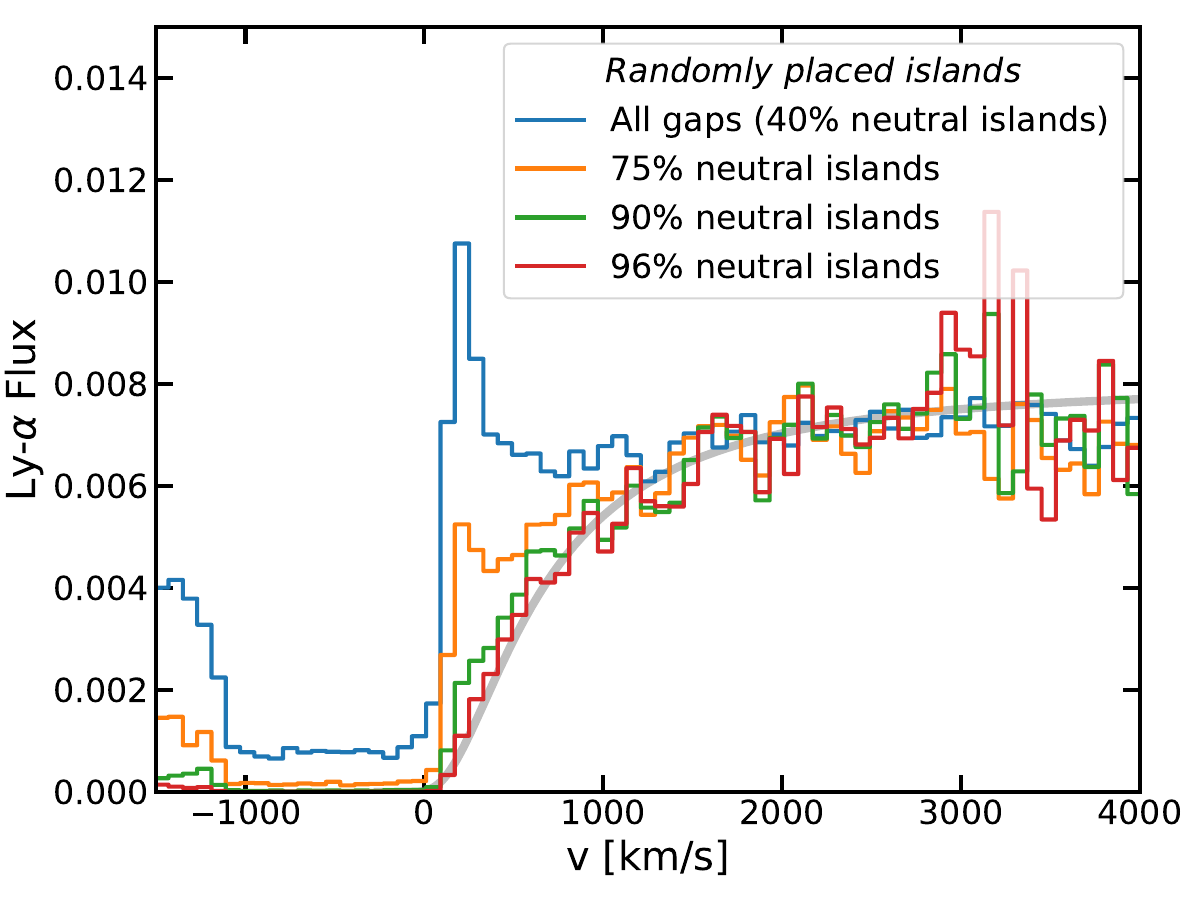}%
\includegraphics[width=0.5\textwidth]{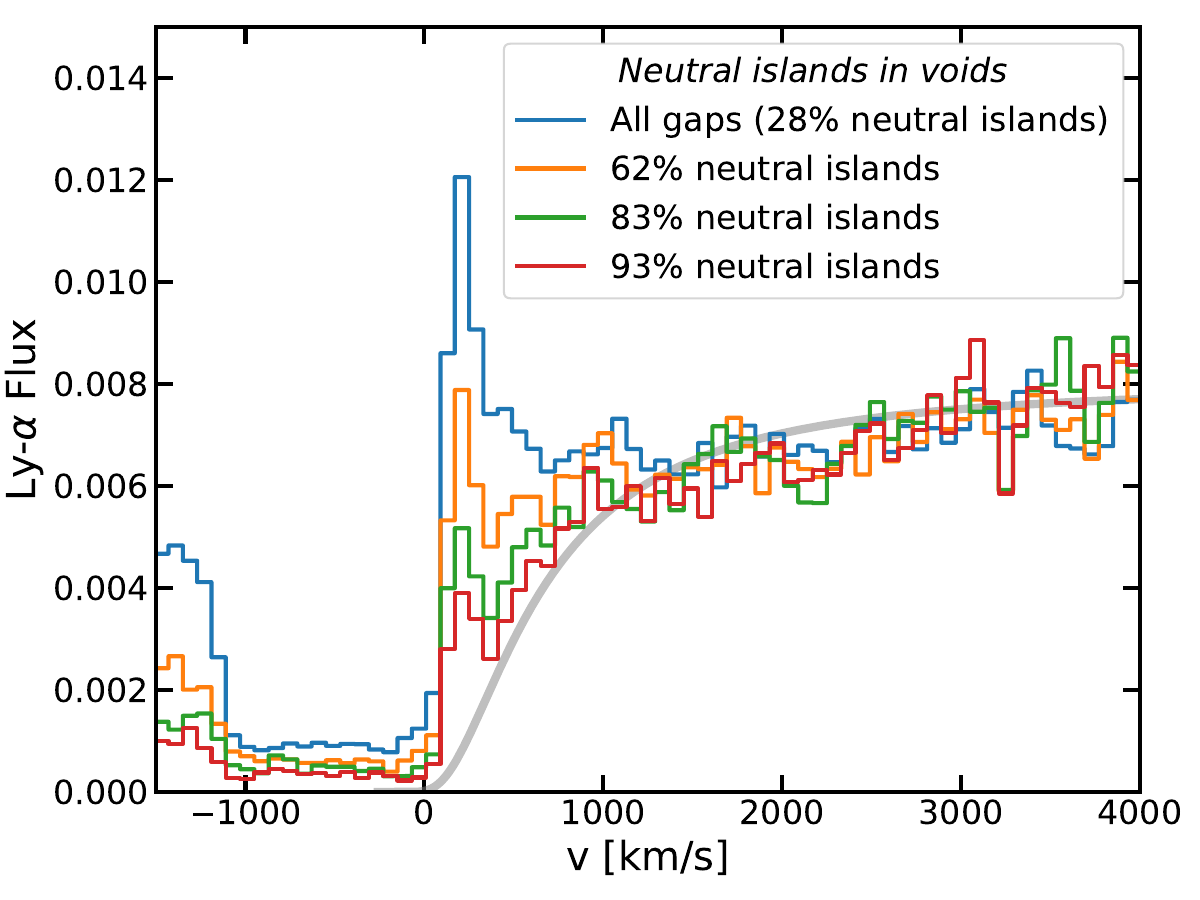}%
\caption{Stacked \Lya\ forest corresponding to the redshift of edges of \Lyb\ dark gaps for the test case with neutral islands added manually, at random location in the left panel and centered on voids in the right panel. Voids are defined as continuous regions of $10 h^{-1}$ cMpc or longer with the mean density of at most 50\% of the cosmic mean. Different lines show tests with progressively larger fractions of neutral islands among all gaps (achieved by lowering the thresholds for gap detection since neutral islands have effectively zero flux). Only when the fraction of neutral islands exceeds 90\% does the damping wing feature become unambiguously detected, and placing neutral islands in the voids makes the damping wing signal \emph{weaker} because large voids are rare}.
\label{fig:l20}
\end{figure*}

Results of this exercise are shown in Figure \ref{fig:gaps}. Neither CROC nor Thesan shows the damping-wing-like feature despite having the number of dark gaps per unit length comparable to the observed frequency, and this result remains robust irrespective of whether observational noise is added to the synthetic spectra as long as the noise level does not exceed the actual noise in observational measurements. In the CROC ``C80C.DC=-1" simulation, reionization ends too early at $z=6.7$ and thus no ``neutral islands" remain at $z=5.9$. In Thesan, reionization proceeds more gradually and is not finished at the end point of the simulation at $z=5.5$. Neutral gas at around the mean cosmic densities still exists in Thesan-1 at $z=5.9$, but apparently, there is not enough of that gas left to produce the damping-wing-like features around the dark gaps. Figure \ref{fig:props} shows the most important properties - the mean neutral fraction vs the mean density and the column density - of dark gaps in the two simulations, where the averaging is done over each gap length. Both simulations have large numbers of dark gaps that are \emph{not} neutral islands. 

In order to explore the discrepancies between simulations and observations better, we manually add $20/h$ cMpc neutral islands to each CROC sightline at random locations (at this size, the actual size of the island becomes irrelevant, the damping wings from islands larger than $20/h$ cMpc do not depend on the island size), as is illustrated in Figure \ref{fig:los}, and repeat the procedure for stacking the dark gaps. With 1000 sightlines, this adds 1000 neutral islands to 1500 dark gaps already present in the original simulation data (i.e.\ with neutral islands making 40\% of all dark gaps). The left panel of Figure \ref{fig:l20} shows the stacked dark gap profiles for this test case - the 40\% neutral island fraction is not enough to exhibit any sign of the damping wing. In order to increase that fraction, we lower the thresholds for the dark gap detection (0.05 for \Lya\ and 0.02 for \Lyb) uniformly by factors of 3, 10, and 30, resulting in the neutral island fraction increasing to 75\%, 90\%, and 96\%, respectively - because without noise the flux inside neutral islands is effectively zero. It is clear from the figure that the spike at the gap edge disappears only when the fraction of neural islands exceeds 75\%, and the fitting function from \citet{YZhu2024} is only reproduced when that fraction is over 95\%. Allowing for a reasonable statistical fluctuation, one can confidently conclude that the damping wing feature only becomes unambiguously detectable when the fraction of gaps that are neutral islands approaches 90\%.

However, as the anonymous referee pointed out, placing neutral islands into random locations may bias this test, as there exists observational evidence that the longest gaps are located in the voids \citep{Christenson2021,Christenson2023}. While presently this evidence remains somewhat controversial \citep{Jin2024,Kashino2025}, such a possibility needs to be explored. To that end, we show in the right panel of Figure \ref{fig:l20} the same test with neutral islands placed only in voids. We define voids as continuous segments of at least $10/h$ cMpc in length that have a mean density of at most 0.5 of the cosmic mean. Such voids do not occur along every line of sight (only along about 60\% of them), and hence the number of thus introduced neutral islands is smaller compared to the random placement case. As a result, the damping wing effect is somewhat weaker. We also tried increasing the minimum void length to the value of the gap length of $20/h$ cMpc. Such voids are so rare that the damping wing effect does not show up at all. We also obtain a very similar result if we place neutral islands in overdense regions rather than in voids - as long as the fraction of the number of lines of sight that contain neutral islands is the same, it does not really matter whether neutral islands are placed in voids or overdense regions.

Finally, given a modest observational sample size, one may wonder if the observational detection of the damping wing is a random statistical fluctuation. In order to check that, we considered the distribution of the damping wing profiles for samples of 24 dark gaps \citep[the sample size of][]{YZhu2024}. In both original simulations, only 1\% of all samples exhibit the damping wing feature similar to the observed one, so it is very unlikely that the disagreement between CROC and Thesan and the observations is a statistical fluke.

\section{Discussion}\label{sec:disc}

The conclusions of this paper would be simple - the simulations fail to reproduce the detection of the damping wing feature in observational data - if one nagging question was not still lingering: where are the ``non-island" dark gaps in the observational data? These are abundant in both CROC and Thesan. In fact, \emph{all} dark gaps in CROC simulations are ``non-island", and their abundance is consistent with observations. Given that the damping wing feature is only observable when neutral islands make up at least around 90\% of all dark gaps, the absence of these ``non-island" gaps like the second one in Figure\ \ref{fig:los} in observations is extremely puzzling. In fact, this was already noticed by \citet[][see their Fig.\ 4e]{YZhu2024}.

This conclusion is not an artifact of simulations not having the correct reionization history. In fact, neither in Thesan nor in CROC do the stacked \Lya\ profiles exhibit the damping wing feature similar to the observed one \emph{at any redshift}, even when reionization is not complete in the simulations. 

One can wonder where the dark gaps in the simulations come from. As guidance for developing physical intuition, we show in the Appendix examples of dark gaps from both simulations. All dark gaps in CROC and the majority of dark gaps in Thesan (see Figure \ref{fig:props}) are ionized. Ionized dark gaps appear when the line of sight happens to run along a large filament or a chain of smaller ones, avoiding deep voids from which large spikes (that terminate gaps) originate. \citet{Gnedin2022} showed that the observed abundance of dark gaps is consistent with the probability of such a chance alignment within the standard large-scale structure theory. The longest dark gap in Thesan does contain several neutral islands, but they all are small and, most importantly, \emph{do not terminate} the gap. The longest gap in Thesan is also terminated in the same way as gaps in CROC and smaller Thesan gaps - by running into a deep enough void. Hence, the lack of such chance alignments in the observations is puzzling.

When dealing with simulations, one should always be comprehensive of the finite numerical resolution and simulation volume. As we mentioned above, CROC and Thesan were carefully designed to (barely) resolve all the structure in the post-reionization IGM. This requirement limits the possible size of the simulation box to about 100 cMpc (for a $2000^3$ simulation). Unfortunately, at present, there are limited opportunities to test whether such a volume is sufficient. In CROC, we have access to smaller, $40 h^{-1}\approx 60$ cMpc boxes. When we repeat the same procedure in these smaller simulations, we find very similar stacked \Lya\ profiles to Figure \ref{fig:gaps}. I.e., we see no box size effect going from 60 cMpc to 120 cMpc boxes, but we are unable to test larger box sizes.

Another surprising feature of the observational claims is the long persistence of neutral islands. An ionization front propagates into a neutral island with the mean neutral hydrogen density $n_H$ with the velocity $v_I$ that satisfies simple photon conservation:
\[
    n_H v_I = c n_I,
\]
where $n_I$ is the incident number density of ionizing photons. The left-hand side is the rate (per unit area) at which neutral hydrogen atoms are ionized by the advancing front, and the right-hand side is the flux of ionizing photons. The latter is easy to estimate from the observationally determined photoionization rate $\Gamma$,
\[
  \Gamma = c \sigma_I n_I,
\]
where $\sigma_I$ is the effective cross-section for an ionizing photon, $\sigma_I=10^{-18}\dim{cm}^2$ if we take the average energy of an ionizing photon to be 25 eV. At $z=5.9$ $\Gamma$ is estimated to be $(1.5-2) \times 10^{-13}\dim{s}^{-1}$ \citep{Gaikwad2023gamma,Davies2024gamma}. Conservatively taking the lower value, we find $c n_I \approx 1.5\times 10^5 \dim{cm}^{-2}\dim{s}^{-1}$ and $v_I\approx 24{,}000$ km/s. At this speed, a 10 cMpc neutral island (a value used by \citet{YZhu2021gaps}) is ionized in 60 Myr, and probably even faster, since such a long sightline is highly likely to cross a cosmic void and thus be on average underdense, plus $\Gamma$ at $z=5.6$ is twice as large as at $z=5.9$. Hence, it is rather surprising that \citet{Spina2024} detected the damping wing features as late as $z=5.6$ (and for significantly shorter gaps). 

\begin{figure}
\centering
\includegraphics[width=\columnwidth]{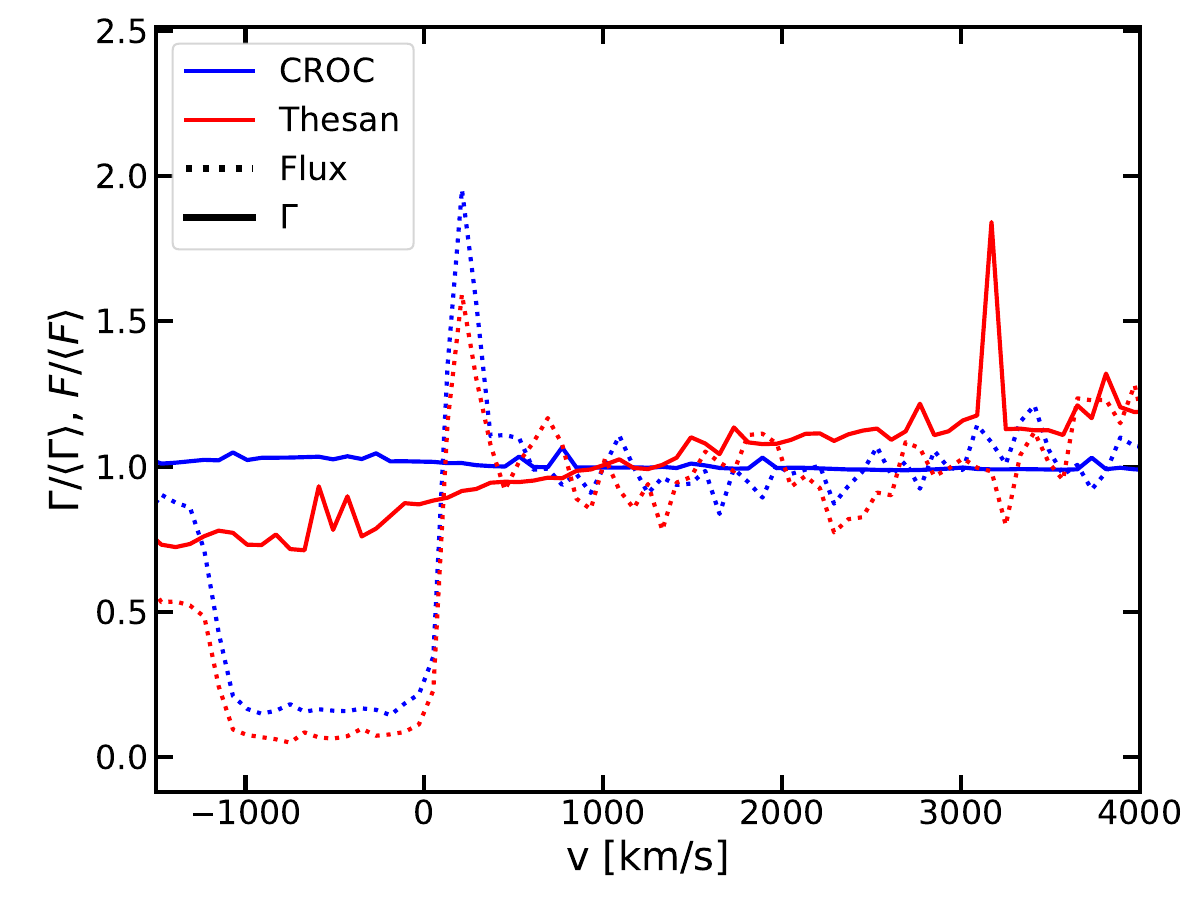}%
\caption{Stacked \Lya\ forest (dotted lines) and the photoionization rate $\Gamma$ (solid lines) corresponding to the redshift of edges of \Lyb\ dark gaps for CROC (blue) and Thesan (red). The two dotted lines are the purple lines from two panels of Figure \ref{fig:gaps} normalized to the mean. CROC shows no systematic variation in the photoionization rate at the edges of the dark gaps, while Thesan does show a small decrease, but that decrease is too small to appreciably extend the lifetimes of neutral islands.}
\label{fig:gamma}
\end{figure}

This argument is so simple that it is hard to get around. Since both $c$ and $\langle n_H\rangle$ are fixed and neutral islands are unlikely to reside in significantly overdense regions ($n_H \la \langle n_H\rangle$), the only possible way out is to reduce $n_I$, i.e.\ to assume that the photoionization rate is suppressed relative to the mean value \emph{everywhere} around neutral islands (Martin Haehnelt, private communication). In order to check whether this is the case in CROC and Thesan, we show in Figure \ref{fig:gamma} the stacked profiles of the \Lya\ forest (the same as shown in Figure \ref{fig:gaps}) as well as stacked profiles of the photoionization rate. In CROC, there is no systematic variation in the photoionization rate at the edges of the dark gaps. In Thesan, there is indeed a slight, $\approx 20\%$ decrease in the photoionization rate in the vicinity of dark gaps, but that decrease is too small to appreciably extend the lifetimes of neutral islands. This discrepancy may also point to the reason for the failure of the simulations: if the simulations underestimate the fluctuations in the photoionization rate (by, for example, not capturing the population of sources correctly, or underestimating the fluctuations in the escape fraction, or not resolving the the full range of Lyman limit systems, and so on and so forth), then they miss a population of long-lived neutral islands, located in the large downward fluctuations of the photoionization rate.

In summary, it appears that the observational detections of damping wing features present serious challenges to the state-of-the-art fully-coupled simulations of reionization.

\acknowledgments

We thank ChatGPT for producing most of the analysis code used in this work and the Thesan team for making their data publicly available. The earlier draft of this manuscript was substantially improved by detailed comments from Yongda Zhu, George Becker, Enrico Garaldi, and Benedetta Spina. This work was supported in part by the NASA Theoretical and Computational Astrophysics Network (TCAN) grant 80NSSC21K0271. This work was supported in part by Fermi Forward Discovery Group, LLC, under Contract No.\ 9243024CSC000002 with the U.S. Department of Energy, Office of Science, Office of High Energy Physics. This work used resources of the Argonne Leadership Computing Facility, which is a DOE Office of Science User Facility supported under Contract DE-AC02-06CH11357. An award of computer time was provided by the Innovative and Novel Computational Impact on Theory and Experiment (INCITE) program. This research is also part of the Blue Waters sustained-petascale computing project, which is supported by the National Science Foundation (awards OCI-0725070 and ACI-1238993) and the state of Illinois. Blue Waters is a joint effort of the University of Illinois at Urbana-Champaign and its National Center for Supercomputing Applications. We also acknowledge the support from the grant NSF PHY-2309135 to the Kavli Institute for Theoretical Physics (KITP) and the support from the University of Chicago’s Research Computing Center.

\end{CJK*}

\bibliographystyle{mnras}
\bibliography{main}

\begin{appendix}


\begin{figure}[H]
\centering
\includegraphics[width=\textwidth]{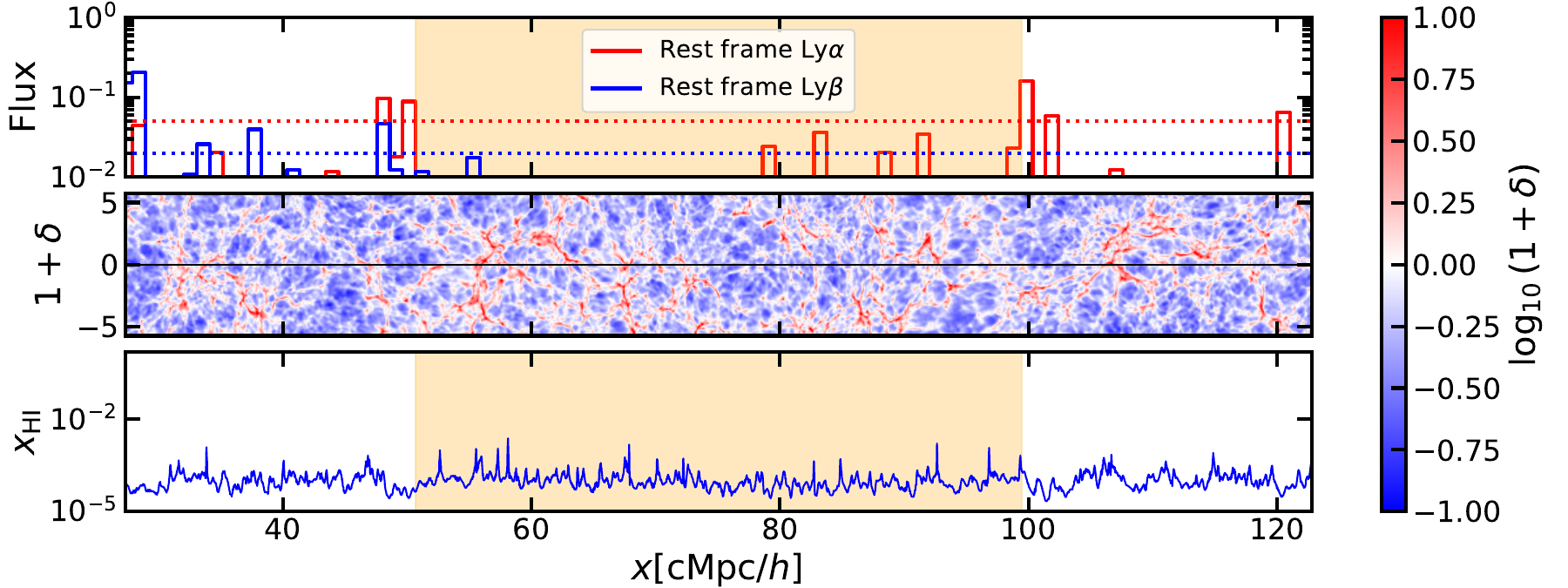}%
\caption{The three panels show (1) the synthetic quasar spectrum (in log scale to show the full dynamic range and binned to 150 km/s resolution as in \citet{YZhu2021gaps}) along one line of sight, (2) the density slice along that line of sight, and (3) the neutral hydrogen fraction along the same line of sight for the longest gap in the CROC simulation (labeled with translucent orange background). Two horizontal dotted lines in the top panel mark the \citet{YZhu2021gaps} threshold for dark gap definition: 0.05 and 0.02 for \Lya\ and \Lyb\ respectively.}
\label{fig:maplos1}
\end{figure}

\begin{figure}[H]
\centering
\includegraphics[width=\textwidth]{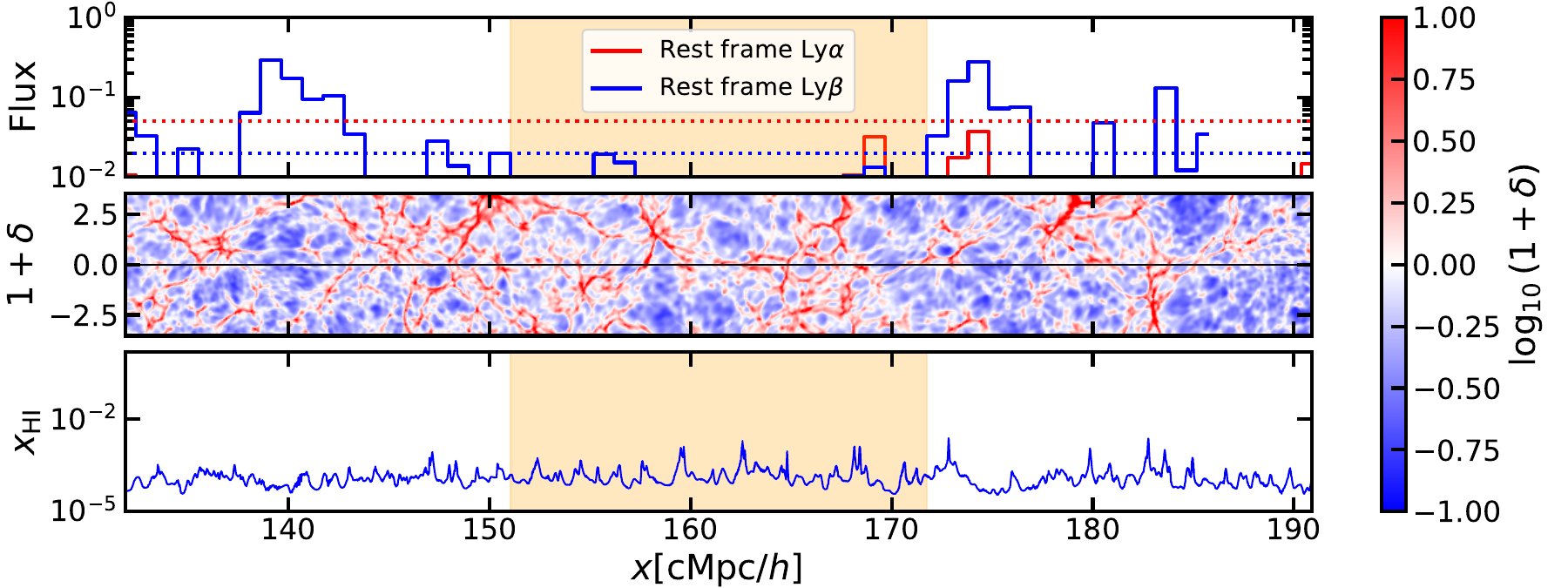}\newline
\includegraphics[width=\textwidth]{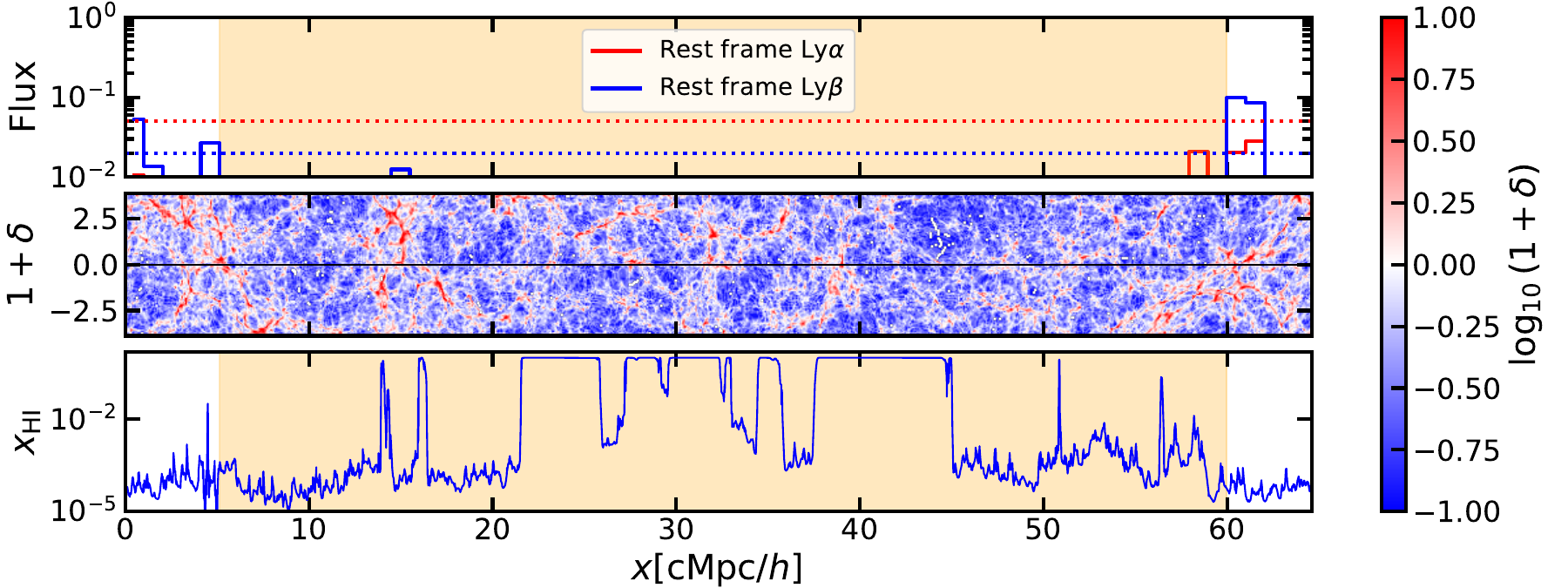}\newline
\includegraphics[width=\textwidth]{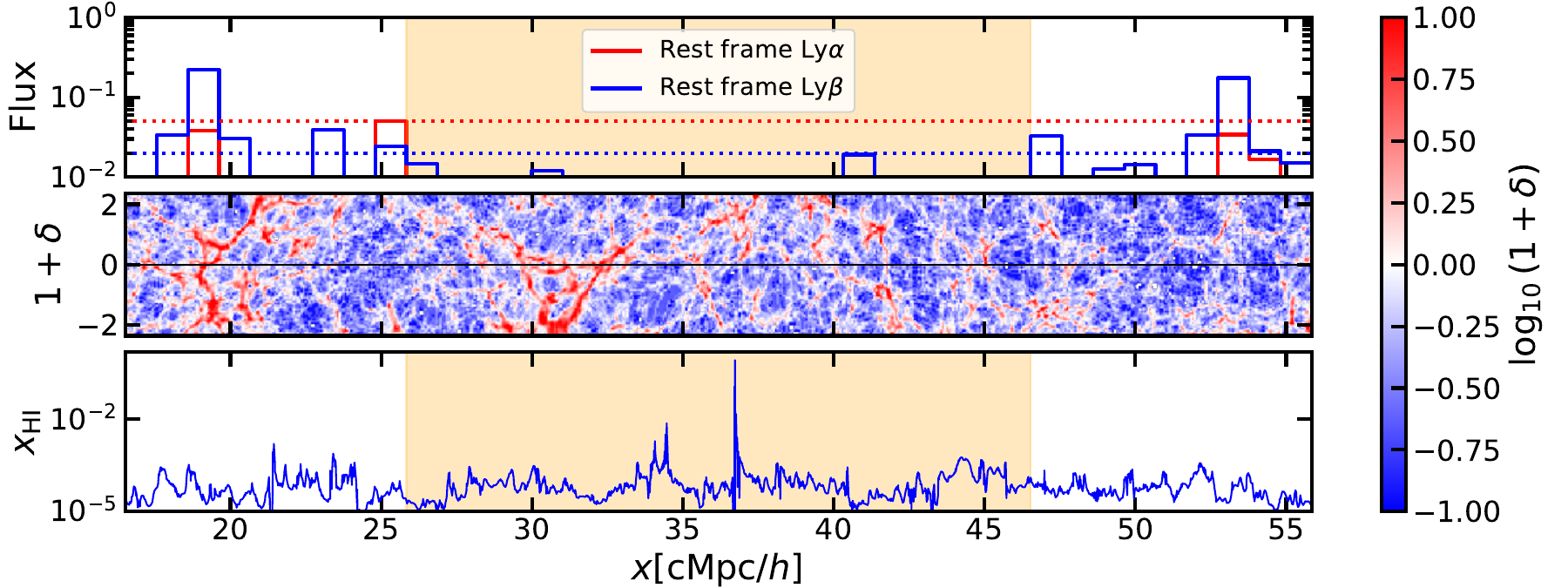}%
\caption{The same as Fig.\ \ref{fig:maplos1} for a randomly chosen CROC gap (top), the longest Thesan gap (middle), and a randomly chosen Thesan gap (bottom).}
\end{figure}

\end{appendix}

\end{document}